\documentclass[aps,prd,twocolumn,superscriptaddress,nofootinbib,floatfix]{revtex4-2}

\usepackage{amsmath,amssymb,bm}
\usepackage{graphicx}
\usepackage{booktabs}
\usepackage{microtype}

\usepackage{xcolor}

\definecolor{linkblue}{RGB}{0,70,160}
\definecolor{revisionblue}{RGB}{0,80,180}

\usepackage[
    colorlinks=true,
    linkcolor=linkblue,   
    citecolor=linkblue,   
    urlcolor=linkblue     
]{hyperref}

\graphicspath{{figures/}}
\newcommand{\dd}{\mathrm{d}}
\newcommand{\GeV}{\mathrm{GeV}}
\newcommand{\TeV}{\mathrm{TeV}}
\newcommand{\MeV}{\mathrm{MeV}}
\newcommand{\keV}{\mathrm{keV}}
\newcommand{\kms}{\mathrm{km\,s^{-1}}}
\newcommand{\erf}{\mathrm{erf}}

\begin{document}

\title{Xenon Isotope Filtering at the Kinematic Edge of Inelastic Dark Matter}

\author{Waqas Ahmed}
\email{waqasmit@hbpu.edu.cn}
\affiliation{Center for Fundamental Physics, School of Artificial Intelligence, Hubei Polytechnic University, Huangshi, China}

\author{Ammara Ahmad}
\email{ammara.ahmad@cern.ch}
\affiliation{Department of Physics, Khalifa University, Abu Dhabi, UAE}

\author{Mansoor Ur Rehman}
\email{mansoor@qau.edu.pk}
\affiliation{Department of Physics, Faculty of Science, Islamic University of Madinah, 42351 Madinah, Saudi Arabia}


\begin{abstract}
Motivated by high-energy xenon recoil searches, we study isotope-dependent endothermic
dark-matter scattering near the upper speed boundary of a truncated Galactic halo.
Small differences among xenon nuclear masses shift the minimum incident speed and can be
strongly amplified when the required velocity approaches the end of the halo distribution.
We formulate the isotope-resolved coherent-contact rate and separate this kinematic filtering
from high-momentum nuclear suppression.  For
$m_\chi=1.1~\TeV$, $\delta=365~\keV$, and $E_R=248~\keV$,
the kinematic-only benchmark gives
$f_{136}^{\rm kin}\simeq0.289$, while after Helm weighting the largest contribution is
$f_{132}\simeq0.346$ and the $^{136}$Xe fraction falls to $\simeq0.094$.
For idealized pure-isotope targets at fixed detector mass, microscopic coupling, and halo,
the integrated $225$--$271~\keV$ responses relative to natural xenon are approximately
$0.404$, $1.235$, and $1.693$ for $^{129}$Xe, $^{132}$Xe, and $^{136}$Xe, respectively.
A single effective-xenon nucleus also fails to reproduce the full isotope sum accurately near
the edge, changing the benchmark window rate by about $8\%$ and producing larger spectral
distortions.  These quantitative results depend on the high-speed halo and on the assumed
nuclear response; the isotope-dependent kinematic support does not.  The calculation is a
theory-level diagnostic rather than an LZ likelihood fit or event-by-event isotope
identification.
\end{abstract}

\maketitle

\section{Introduction}

The LUX-ZEPLIN (LZ) Collaboration has extended its nuclear-recoil search to energies of
approximately $270~\keV$ and reported one event, LZ230616, with characteristics consistent
with a nuclear recoil at
\begin{equation}
E_R=248\pm23\,(\mathrm{stat})\pm23\,(\mathrm{sys})~\keV
\end{equation}
in a $2.84$ tonne-year exposure \cite{LZ2026}.  The largest local significance among the
interaction hypotheses tested by LZ is $3.4\sigma$, while the global significance after the
look-elsewhere effect is $2.6\sigma$ \cite{LZ2026}.  The event therefore does not constitute
evidence for dark matter.  Its unusually large recoil energy is nevertheless a useful
laboratory for interactions with a kinematic threshold or strong momentum dependence.

Endothermic dark matter provides a simple example \cite{TuckerSmith2001,TuckerSmith2005}.
A halo particle $\chi_1$ scatters into a slightly heavier state $\chi_2$,
\begin{equation}
\chi_1+N\rightarrow\chi_2+N,
\qquad
\delta\equiv m_{\chi_2}-m_{\chi_1}>0.
\label{eq:process}
\end{equation}
A fraction of the incoming kinetic energy is then spent on the mass splitting.  For
TeV-scale dark matter and $\delta$ of a few hundred keV, the required speed can approach the
largest velocities available in the laboratory-frame halo, removing most of the low-speed
population and moving spectral support to high recoil energy.

The LZ event has generated a broad phenomenological literature.  
Endothermic explanations
include generic inelastic and pseudo-Dirac scenarios
\cite{Su2026,DiMauro2026,McCabe2026,Yang2026,Fan2026,Visinelli2026}, Higgsino and supersymmetric models
\cite{Freese2026,Cheung2026,Bisal2026}, inert- and singlet-doublet models
\cite{WangXiao2026,Borah2026}, gauged Abelian models
\cite{OkadaSeto2026,KumarPrajapati2026,QiSun2026}, light-mediator and transition-dipole
constructions \cite{Yamashita2026,Zhu2026,Yuan2026,He2026}, and extra-dimensional scenarios
\cite{AhmedLeontaris2026,LeeRandall2026}.  Exothermic scattering
\cite{DentNewstead2026,DeLima2026,BaerBarger2026}, absorption
\cite{LouLu2026}, neutrino up-scattering \cite{JeesunMajumdar2026}, boosted populations
\cite{Alhazmi2026,Kannike2026,Heikinheimo2026,Liang2026}, neutron disappearance
\cite{AghaieStrumia2026}, and nuclear-response effects \cite{Khan2026} provide qualitatively
different alternatives.  Momentum-dependent elastic and additional inelastic constructions
have also been explored \cite{Unwin2026,Das2026}, while a purely elastic-neutrino
interpretation is strongly constrained \cite{Chattaraj2026}.

{\color{black}
Several additional recent constructions further illustrate the diversity of particle-physics
interpretations of LZ230616.  These include an inelastic self-interacting dark-matter
realization based on a modular $A_4$ Dirac inverse-seesaw framework
\cite{Das2026}, a generalized chiral $U(1)_{B-L}$ model with inelastic scalar
dark matter \cite{KumarPrajapati2026}, and an endothermic light-dark-photon
scenario \cite{Zhu2026}.  A complementary electroweak interpretation based on
singlet--doublet Majorana dark matter with a vector-like lepton sector instead
uses a hard spin-dependent recoil spectrum and is independently testable
through solar-neutrino searches \cite{ElahiSchwaller2026}.
}

For the present work, four comparisons are especially relevant.
Ref.~\cite{DiMauro2026} emphasizes the kinematic-edge nature of the
high-recoil event, while Ref.~\cite{McCabe2026} shows that endothermic
interpretations can be particularly sensitive to the seasonal motion of the Earth. Ref.~\cite{Khan2026} discusses high-$q$ nuclear interference as an additional source of spectral structure. In addition,
Ref.~\cite{XENON1T2024} treats the xenon isotopes separately and combines their individual recoil spectra according to their natural abundances. Our purpose is therefore not to introduce isotope-resolved scattering.
Instead, we investigate whether the \emph{relative isotope weights}
themselves become strongly energy dependent and rapidly varying when an endothermic signal lies close to a kinematic boundary.

{\color{black}
Solar capture provides an important model-dependent complement to this terrestrial
kinematic picture.  For a canonical full-density thermal Higgsino with
$m_\chi\simeq1.08~\TeV$, recent analyses that map the IceCube solar-neutrino
limits onto endothermic capture require a substantially larger neutral-state
splitting, approximately
\begin{equation}
\delta_{\rm IC}^{\tilde H}\gtrsim (0.51\text{--}0.57)~\MeV ,
\label{eq:icecubecontext}
\end{equation}
depending on the treatment of post-capture evolution
\cite{PospelovRamani2026,DiMauroShaikh2026,IceCube2025}.
This should not be interpreted as a model-independent lower bound on the
splitting: the translation of a solar-neutrino limit into a constraint on
$\delta$ depends on the scattering operator, capture rate, annihilation
channels, and the equilibration of the captured population.  We therefore use
IceCube only as an external model-specific consistency test and keep the
isotope-filtering analysis itself interaction-transparent.
}

Natural xenon contains nine naturally occurring, long-lived isotopes.  Their masses differ
only at the percent level, but the endothermic minimum speed depends nonlinearly on the target
mass.  Deep inside the velocity distribution these mass differences cause only mild changes.
Near the upper support boundary, however, the same small shifts are amplified by the rapidly
vanishing mean inverse speed.  We refer to this edge-enhanced redistribution as
\emph{xenon isotope filtering}.

A second effect is numerically comparable.  A recoil near $250~\keV$ in xenon corresponds to
$q\simeq0.25~\GeV$, where coherent nuclear responses are already strongly suppressed and can
vary rapidly with isotope.  The benchmark therefore contains a competition:
\begin{equation}
\begin{gathered}
\text{heavier isotopes}
\Rightarrow
v_{\min}^{A}\downarrow
\Rightarrow
\eta(v_{\min}^{A})\uparrow,
\\[1mm]
\text{while at large }q,\qquad
F_A^2(q)\downarrow .
\end{gathered}
\label{eq:competition}
\end{equation}
The first trend is kinematic; the second is nuclear.  Keeping them separate is central to the analysis below.

We work with a deliberately transparent baseline: an
isospin-conserving coherent contact interaction, a truncated
Standard Halo Model (SHM), and a Helm nuclear form factor.
Our purpose is to isolate the isotope-dependent kinematics and
its interplay with the high-$q$ nuclear response rather than to
perform a detector-level fit.  In particular, we do not carry out
an official LZ likelihood recast, which would additionally require
the detector response, backgrounds, sidebands, efficiencies, and
the full likelihood construction.

The paper is organized as follows.  In
Sec.~\ref{sec:kinematics} we derive the isotope-dependent
endothermic kinematics and recoil boundaries.  In
Sec.~\ref{sec:rate} we construct the isotope-resolved scattering
rate and separate the kinematic and nuclear contributions.
Section~\ref{sec:numerics} specifies the halo model, xenon inputs,
and benchmark setup.  The main isotope-filtering results are
presented in Sec.~\ref{sec:results}, while
Sec.~\ref{sec:halo} examines the dependence on the high-speed halo.
Sections~\ref{sec:nuclear} and \ref{sec:experimental} discuss the
nuclear-response and detector-level limitations, respectively.
{\color{black}Section~\ref{sec:icecube} places the benchmark in the context of
model-dependent solar-capture constraints from IceCube.}
We compare the mechanism with related LZ interpretations in
Sec.~\ref{sec:relation}, summarize the main control checks in
Sec.~\ref{sec:discussion}, and conclude in
Sec.~\ref{sec:conclusions}.

\section{Isotope-resolved endothermic kinematics}
\label{sec:kinematics}

Throughout the analytical formulae we use $\hbar=c=1$ and identify
$m_\chi\equiv m_{\chi_1}$.  The speed of light is restored when velocities are quoted in
$\kms$.  The symbol $A$ denotes the isotope mass number.

\subsection{Minimum speed}

For nonrelativistic scattering from a target nucleus of mass $m_A$, the minimum incoming
speed required to produce recoil energy $E_R$ is
\begin{equation}
v_{\min}^A(E_R)
=
\frac{m_A E_R/\mu_{\chi A}+\delta}
{\sqrt{2m_AE_R}},
\label{eq:vmin}
\end{equation}
where
\begin{equation}
\mu_{\chi A}
=
\frac{m_\chi m_A}{m_\chi+m_A}
\label{eq:muA}
\end{equation}
is the dark-matter--nucleus reduced mass.  The minimum of this function occurs at
\begin{equation}
E_{R,A}^{\star}
=
\frac{\delta\,\mu_{\chi A}}{m_A},
\qquad
v_{\min,A}^{\star}
=
\sqrt{\frac{2\delta}{\mu_{\chi A}}}.
\label{eq:edge}
\end{equation}

The familiar statement that a heavier isotope has a lower threshold is not universal for all
kinematics.  At fixed $E_R$ and $\delta$, writing $m$ for the target mass gives
\begin{equation}
v_{\min}(m)
=
\frac{(E_R+\delta)+E_Rm/m_\chi}
{\sqrt{2mE_R}},
\end{equation}
and hence
\begin{equation}
\frac{\partial\ln v_{\min}}{\partial\ln m}
=
\frac{E_Rm/m_\chi-(E_R+\delta)}
{2[E_Rm/m_\chi+(E_R+\delta)]}.
\label{eq:massderivative}
\end{equation}
Thus $v_{\min}$ decreases with increasing target mass when
\begin{equation}
m<m_\chi\left(1+\frac{\delta}{E_R}\right).
\label{eq:masscondition}
\end{equation}
This condition is comfortably satisfied by xenon at the TeV-scale benchmark studied below.

\subsection{Isotope-dependent recoil window}

For a particle of fixed incident speed $v$, scattering is possible only if
\begin{equation}
v^2\geq \frac{2\delta}{\mu_{\chi A}}.
\end{equation}
When this condition is satisfied, the allowed recoil energies are
\begin{equation}
E_{R,A}^{\pm}(v)
=
\frac{\mu_{\chi A}^2 v^2}{2m_A}
\left[
1\pm
\sqrt{1-\frac{2\delta}{\mu_{\chi A}v^2}}
\right]^2 .
\label{eq:endpoints}
\end{equation}
The isotope dependence therefore enters through the entire allowed recoil interval, not only
through the value of $v_{\min}$.

For a sharply truncated halo the maximum laboratory-frame speed is
\begin{equation}
v_{\max}=v_{\rm esc}+v_E,
\label{eq:vmax}
\end{equation}
within the adopted SHM.  A recoil is supported only when
\begin{equation}
E_{R,A}^{-}(v_{\max})
<
E_R
<
E_{R,A}^{+}(v_{\max}).
\label{eq:allowedband}
\end{equation}
The hard boundary in Eq.~\eqref{eq:vmax} is a property of the sharply truncated halo model,
not a universal physical cutoff.

\section{Isotope-resolved rate and nuclear response}
\label{sec:rate}

To expose the target-mass effect with minimal particle-physics assumptions, we use an
isospin-conserving coherent contact interaction,
\begin{equation}
\frac{\dd\sigma_A}{\dd E_R}
=
\frac{m_A\,\sigma_n A^2}
{2\mu_{\chi n}^2v^2}
F_A^2(q),
\label{eq:dsigma}
\end{equation}
where $\sigma_n$ is the reference dark-matter--nucleon cross section and
\begin{equation}
\mu_{\chi n}
=
\frac{m_\chi m_n}{m_\chi+m_n}.
\end{equation}
The momentum transfer is
\begin{equation}
q=\sqrt{2m_AE_R}.
\label{eq:qdef}
\end{equation}

Let $\xi_A$ be the natural \emph{atom fraction}, with $\sum_A\xi_A=1$, and define
\begin{equation}
\overline m_{\rm Xe}
=
\sum_A \xi_A m_A .
\label{eq:mbar}
\end{equation}
There are $\xi_A/\overline m_{\rm Xe}$ nuclei of isotope $A$ per unit detector mass.  The
differential recoil rate is therefore
\begin{align}
\frac{\dd R}{\dd E_R}
&=
\frac{\rho_\chi}{m_\chi}
\frac{1}{\overline m_{\rm Xe}}
\sum_A\xi_A
\int_{v>v_{\min}^A}
\dd^3v\,
f_E(\bm v)\,v\,
\frac{\dd\sigma_A}{\dd E_R}
\nonumber\\[1mm]
&=
\frac{\rho_\chi\sigma_n}
{2m_\chi\mu_{\chi n}^2\overline m_{\rm Xe}}
\sum_A
\xi_A\,m_AA^2F_A^2(q)\,
\eta\!\left(v_{\min}^A\right),
\label{eq:rate}
\end{align}
where
\begin{equation}
\eta(v_{\min})
=
\int_{v>v_{\min}}
\frac{\dd^3v}{v}\,
f_E(\bm v)
\label{eq:eta}
\end{equation}
is the mean inverse speed.  Because $\xi_A$ is an atom fraction, the isotope kernel contains
one explicit power of $m_A$.  An equivalent mass-fraction convention gives the same total
rate when used consistently.

For the baseline nuclear response we use the Helm form factor \cite{LewinSmith1996},
\begin{equation}
F_A(q)
=
3\frac{j_1(qr_{n,A})}{qr_{n,A}}
\exp\left[-\frac{(qs)^2}{2}\right],
\label{eq:helm}
\end{equation}
where $j_1$ is the spherical Bessel function,
\begin{equation}
r_{n,A}^2
=
c_A^2+\frac{7\pi^2a^2}{3}-5s^2,
\qquad
c_A=1.23A^{1/3}-0.60~{\rm fm},
\label{eq:helmpars}
\end{equation}
and
\begin{equation}
a=0.52~{\rm fm},
\qquad
s=0.90~{\rm fm}.
\end{equation}
Numerically, $q$ is converted to ${\rm fm}^{-1}$ using
$\hbar c=0.1973269804~\GeV\,{\rm fm}$.

The Helm prescription is intentionally a proof-of-principle baseline.  At
$q\simeq0.25~\GeV$, xenon nuclear responses can contain substantial isotope dependence and
interference structure \cite{Fitzpatrick2013,Anand2014,Khan2026}.  The kinematic calculation
below is therefore kept separate from the nuclear model.

For each isotope we define
\begin{equation}
K_A(E_R)
=
\xi_A m_AA^2F_A^2(q)\,
\eta\!\left(v_{\min}^A(E_R)\right).
\label{eq:kernel}
\end{equation}
When the total response is nonzero, the normalized full fraction is
\begin{equation}
f_A(E_R)
=
\frac{K_A(E_R)}
{\sum_BK_B(E_R)}.
\label{eq:fraction}
\end{equation}
Where $\sum_BK_B=0$, $f_A$ is undefined and is masked in all numerical plots.

To isolate the halo effect we define
\begin{equation}
f_A^{\rm kin}(E_R)
=
\frac{
\xi_Am_AA^2\,\eta(v_{\min}^A)
}{
\sum_B\xi_Bm_BB^2\,\eta(v_{\min}^B)
},
\label{eq:fkin}
\end{equation}
which sets $F_A^2\to1$.  A complementary nuclear-only fraction is
\begin{equation}
f_A^{\rm nuc}(E_R)
=
\frac{
\xi_Am_AA^2F_A^2(q)
}{
\sum_B\xi_Bm_BB^2F_B^2(q)
}.
\label{eq:fnuc}
\end{equation}
Finally, when $^{132}$Xe has nonzero halo support, we use
\begin{equation}
H_A(E_R)
=
\frac{
\eta[v_{\min}^A(E_R)]
}{
\eta[v_{\min}^{132}(E_R)]
}
\label{eq:haloenhancement}
\end{equation}
as a purely kinematic ratio.

\section{Halo model and numerical setup}
\label{sec:numerics}

We use a truncated Maxwellian SHM \cite{LewinSmith1996}, boosted into the Earth frame.  The
representative benchmark values are
\begin{equation}
\begin{aligned}
v_0&=238~\kms,
&
v_{\rm esc}&=544~\kms,
\\
v_E&=250.5~\kms.
\end{aligned}
\label{eq:halo}
\end{equation}
The last value is treated as a fixed representative laboratory speed, not as an
exposure-weighted annual average.  Time-dependent predictions require $v_E(t)$; the strong
seasonal sensitivity of endothermic interpretations has been emphasized in
Ref.~\cite{McCabe2026}.  The $v_{\rm esc}$ variation in Sec.~\ref{sec:halo} is a sensitivity
exercise, not a statistical confidence interval.

For
\begin{equation}
z=\frac{v_{\rm esc}}{v_0},
\end{equation}
the truncation normalization is
\begin{equation}
N_{\rm esc}
=
\erf(z)
-
\frac{2z}{\sqrt{\pi}}e^{-z^2}.
\label{eq:Nesc}
\end{equation}
The analytic $\eta(v_{\min})$ used in the calculation is given in
Appendix~\ref{app:eta}.  The implementation uses a stable series close to
$v_{\min}=v_{\rm esc}+v_E$ to avoid cancellation.

Natural atom fractions and neutral-atom isotope masses are taken from the National Institute
of Standards and Technology (NIST) \cite{NISTXe}.  In the scattering kinematics we use the
consistent approximate nuclear-mass convention
\begin{equation}
m_A
=
m_A^{\rm atom}
-
54\,m_e,
\qquad
m_e=0.51099895~\MeV.
\label{eq:massconv}
\end{equation}
The much smaller total electronic binding-energy correction is neglected.  Using the neutral
atomic masses directly would shift the benchmark $v_{\min}$ values by less than
$0.1~\kms$ and does not change the isotope-support ordering reported below.  All figures,
tables, and quoted benchmark numbers use Eq.~\eqref{eq:massconv}.  The complete numerical
input table is given in Appendix~\ref{app:isotopes}.

\subsection{Reference benchmark}

Our reference point is
\begin{equation}
m_\chi=1.1~\TeV,
\qquad
\delta=365~\keV.
\label{eq:benchmark}
\end{equation}
{\color{black}
The benchmark in Eq.~\eqref{eq:benchmark} should therefore be regarded as a
kinematic benchmark for the isotope-filtering mechanism.  If it were
identified specifically with a canonical thermal Higgsino, its
$\delta=365~\keV$ splitting would lie below the solar-capture requirement in
Eq.~\eqref{eq:icecubecontext} and would be excluded by that model-specific
IceCube interpretation.  This does not exclude a generic endothermic contact
interaction with different capture or annihilation dynamics.
}

It is not a best fit to the LZ data.  It is deliberately chosen close to the high-speed edge
to expose the mechanism.  With Eq.~\eqref{eq:halo},
\begin{equation}
v_{\max}=794.5~\kms.
\end{equation}

Table~\ref{tab:benchmark} gives the exact benchmark values used by the final notebook.
At $248~\keV$, $^{124}$Xe and $^{126}$Xe are both outside the sharp halo support.
The latter is close to the boundary,
$v_{\min}^{126}=794.91~\kms$, but still exceeds $v_{\max}$; it turns on only above
$E_R^-(v_{\max})\simeq249.83~\keV$ and therefore contributes to the upper part of the
illustrative recoil window.

\begin{table}[t]
\caption{\textbf{Benchmark kinematics at $E_R=248~\keV$.}
The masses follow Eq.~\eqref{eq:massconv}.  The last column is the Helm baseline.}
\label{tab:benchmark}
\centering
\small
\resizebox{\columnwidth}{!}{
\begin{tabular}{cccc}
\toprule
Isotope &
$v_{\min}$ [km/s] &
$E_R^-(v_{\max})$ [keV] &
$F_A^2$\\
\midrule
$^{124}$Xe & 800.77 & 287.74 & $5.80\times10^{-4}$\\
$^{126}$Xe & 794.91 & 249.83 & $4.36\times10^{-4}$\\
$^{128}$Xe & 789.19 & 228.19 & $3.14\times10^{-4}$\\
$^{129}$Xe & 786.37 & 219.74 & $2.61\times10^{-4}$\\
$^{130}$Xe & 783.60 & 212.33 & $2.14\times10^{-4}$\\
$^{131}$Xe & 780.85 & 205.68 & $1.72\times10^{-4}$\\
$^{132}$Xe & 778.15 & 199.67 & $1.35\times10^{-4}$\\
$^{134}$Xe & 772.81 & 189.10 & $7.61\times10^{-5}$\\
$^{136}$Xe & 767.60 & 180.02 & $3.53\times10^{-5}$\\
\bottomrule
\end{tabular}}
\end{table}

\section{Results: xenon isotope filtering}
\label{sec:results}

\begin{figure}[t]
\centering
\includegraphics[width=\columnwidth]{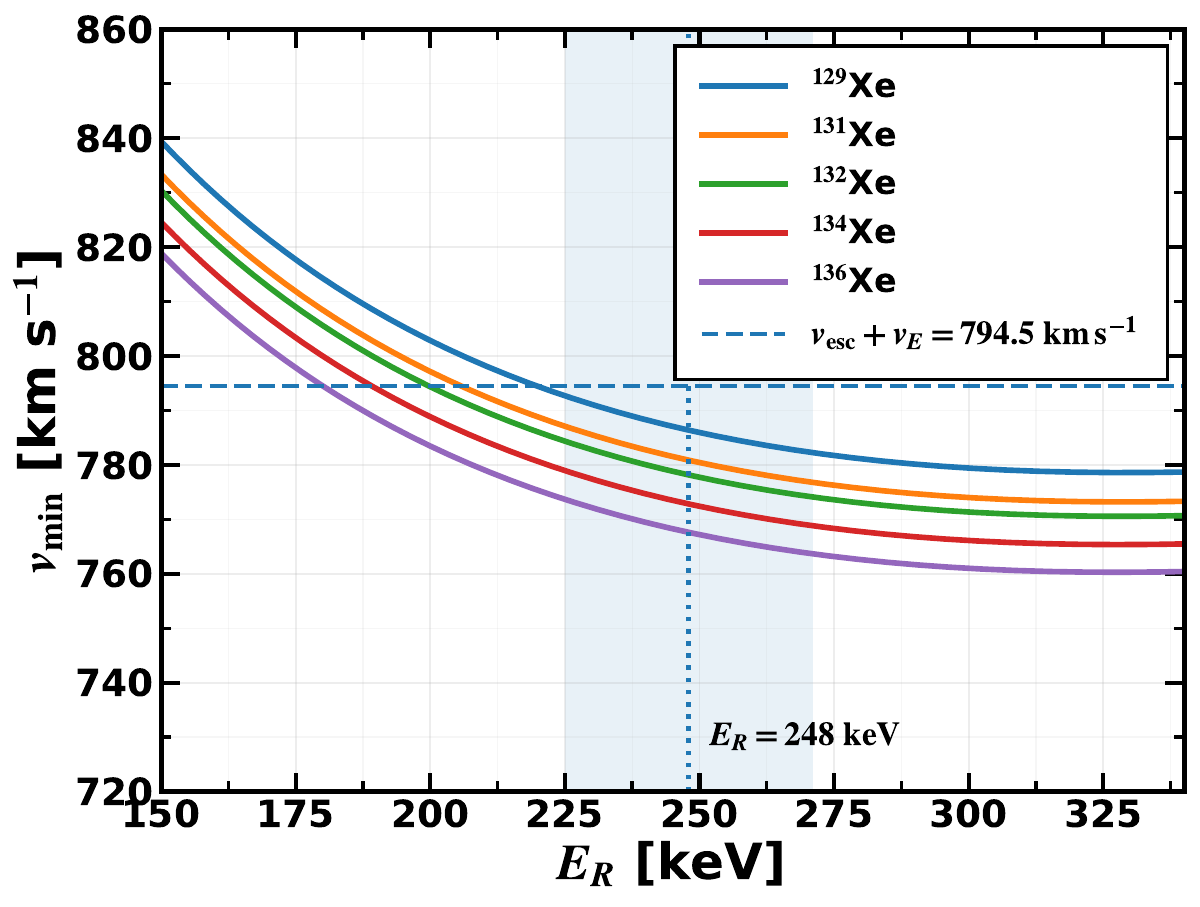}
\caption{\textbf{Minimum incoming speed for the dominant xenon isotopes.}
Curves show $v_{\min}^A(E_R)$ for the benchmark in Eq.~\eqref{eq:benchmark}.
The horizontal dashed line is $v_{\rm esc}+v_E=794.5~\kms$.
The vertical dotted line marks $248~\keV$ and the shaded band denotes the illustrative
$225$--$271~\keV$ interval.}
\label{fig:vmin}
\end{figure}

Figure~\ref{fig:vmin} shows the minimum speed for the five abundant heavy isotopes.  On an
ordinary velocity scale the curves are close, but the relevant quantity is their distance from
$v_{\max}$.  Near $248~\keV$, shifts of only a few $\kms$ sample very different amounts of
the remaining high-speed phase space.

The rare light isotopes are not drawn in Fig.~\ref{fig:vmin}, but Table~\ref{tab:benchmark}
makes their role explicit: $^{124}$Xe has no support anywhere in the
$225$--$271~\keV$ window, while $^{126}$Xe becomes accessible only near its upper portion.
The abundant isotopes $^{129}$Xe--$^{136}$Xe remain supported at $248~\keV$ but probe
significantly different values of $\eta$.

\begin{figure*}[t]
\centering
\includegraphics[width=0.96\textwidth]{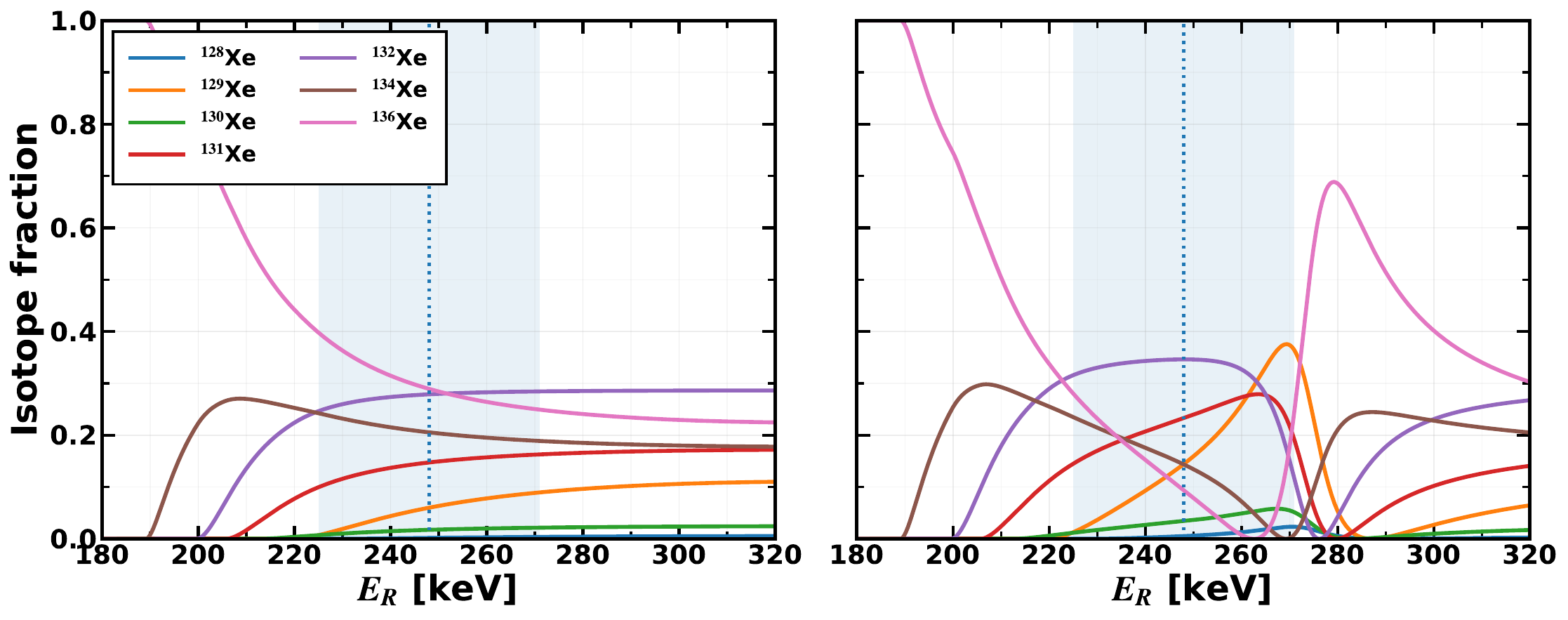}
\caption{\textbf{Decomposition of xenon isotope filtering.}
Left: kinematic-only fractions $f_A^{\rm kin}$ with $F_A^2=1$.
Right: full fractions $f_A$ including the Helm form factor.
All nine isotopes are included in the normalization; isotopes with natural abundance above
approximately $1\%$ are shown.  The shaded band is $225$--$271~\keV$ and the dotted line is
$248~\keV$.}
\label{fig:decomposition}
\end{figure*}

Figure~\ref{fig:decomposition} compares the isotope fractions with and without Helm
weighting, using identical masses, abundances, and halo parameters.

At $248~\keV$ the kinematic-only fractions are
\begin{equation}
f_{136}^{\rm kin}=0.289,
\qquad
f_{132}^{\rm kin}=0.279,
\qquad
f_{134}^{\rm kin}=0.205,
\label{eq:kinfractions}
\end{equation}
while the full Helm-weighted fractions include
\begin{equation}
\begin{aligned}
f_{132}&=0.346,\qquad f_{131}=0.233,\\
f_{129}&=0.144,\qquad f_{136}=0.094.
\end{aligned}
\label{eq:fullfractions}
\end{equation}
The edge therefore favors the heavy end of the isotope distribution, but the high-$q$ Helm
suppression substantially reshuffles the hierarchy.  These numerical fractions are
halo- and nuclear-model dependent; the existence of isotope-dependent thresholds is not.

The sharp structures in the right panel near Helm minima are normalized-fraction effects.
They should be interpreted together with the total response, which is shown separately in
the numerical notebook and is not large at every fraction spike.

We next integrate over the illustrative theory interval
\begin{equation}
225~\keV<E_R<271~\keV.
\label{eq:window}
\end{equation}
This is not the full LZ analysis window and is not a confidence interval; it is simply a band
centered on the reported recoil scale.  We define
\begin{equation}
I_A
=
\int_{225\,\keV}^{271\,\keV}
\dd E_R\,\omega_A(E_R),
\label{eq:IA}
\end{equation}
with
\begin{equation}
\omega_A^{\rm kin}
=
\xi_Am_AA^2\eta(v_{\min}^A),
\qquad
\omega_A^{\rm full}=K_A.
\end{equation}

\begin{figure*}[t]
\centering
\includegraphics[width=0.96\textwidth]{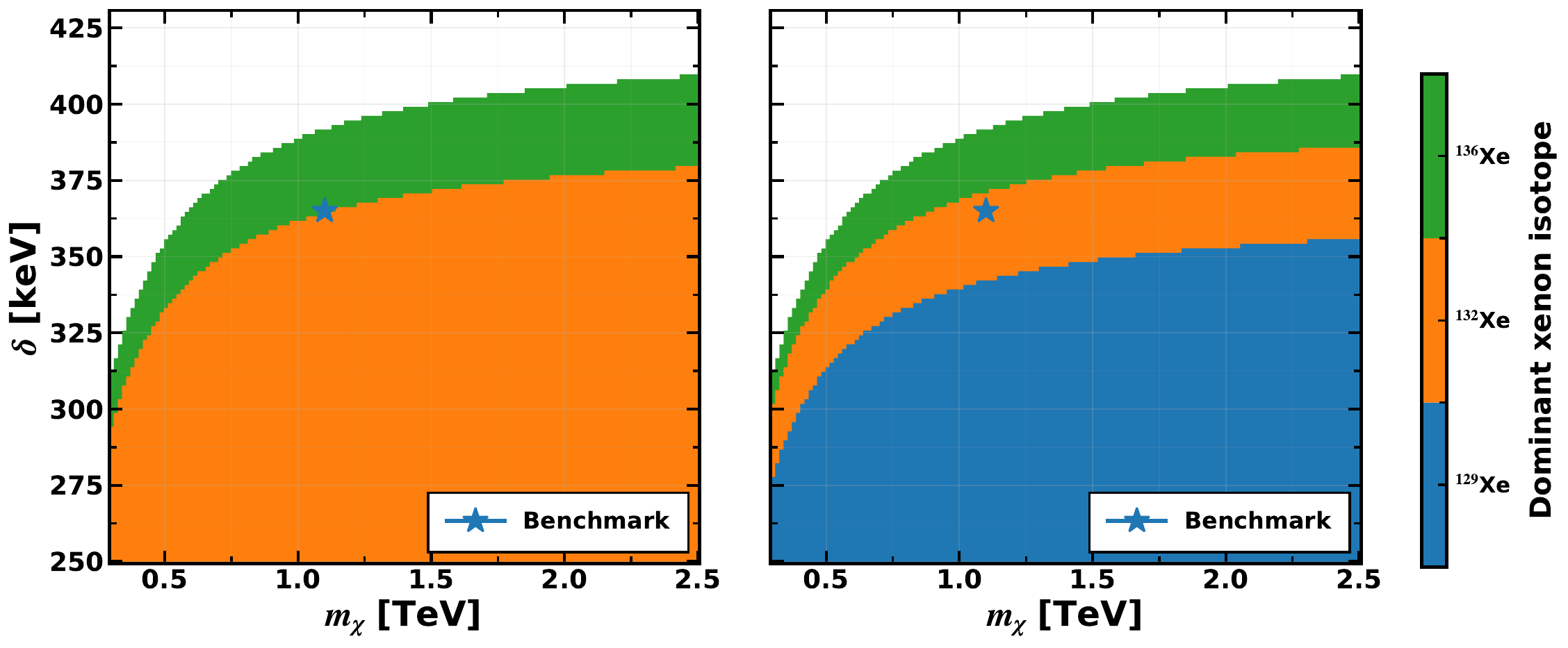}
\caption{\textbf{Dominant xenon isotope in the $(m_\chi,\delta)$ plane.}
Left: kinematic-only integrated response.
Right: Helm-weighted response.
The scan uses $m_\chi=0.3$--$2.5~\TeV$ (141 points),
$\delta=250$--$430~\keV$ (121 points), and 120 recoil-energy samples in
$225$--$271~\keV$ with trapezoidal integration.
The star marks Eq.~\eqref{eq:benchmark}; regions with zero support are left blank.
A common discrete isotope scale is used in both panels.}
\label{fig:dominant}
\end{figure*}

The difference between the two panels is itself a result.  The left panel is controlled by
target masses and the chosen halo function.  The right panel is a prediction for the
specified coherent-contact/Helm baseline and can change for other operators.

For a target composition $X$, define
\begin{equation}
\overline m_X
=
\sum_A\xi_A^{(X)}m_A
\end{equation}
and
\begin{equation}
C
=
\frac{\rho_\chi\sigma_n}
{2m_\chi\mu_{\chi n}^2}.
\end{equation}
The rate per unit detector mass is then
\begin{equation}
\frac{\dd R_X}{\dd E_R}
=
\frac{C}{\overline m_X}
\sum_A
\xi_A^{(X)}m_AA^2F_A^2(q)
\eta(v_{\min}^A).
\label{eq:compositionrate}
\end{equation}
The target-dependent factor $1/\overline m_X$ is essential when different isotope
compositions are compared at equal detector mass.

For spectral shape we use
\begin{equation}
S_X(E_R)
=
\frac{
\dd R_X/\dd E_R
}{
\displaystyle
\int_{180\,\keV}^{320\,\keV}
\dd E_R\,(\dd R_X/\dd E_R)
},
\label{eq:shape}
\end{equation}
while the benchmark rate ratio is
\begin{equation}
R_{X/{\rm nat}}
=
\frac{
\displaystyle
\int_{225\,\keV}^{271\,\keV}
\dd E_R\,(\dd R_X/\dd E_R)
}{
\displaystyle
\int_{225\,\keV}^{271\,\keV}
\dd E_R\,(\dd R_{\rm nat}/\dd E_R)
}.
\label{eq:compositionratio}
\end{equation}
Equation~\eqref{eq:compositionratio} assumes the same detector mass--time exposure, halo,
and microscopic coupling.  It is a true-recoil theory ratio with unit efficiency, not an
accepted-event forecast.

\begin{figure*}[t]
\centering
\includegraphics[width=0.96\textwidth]{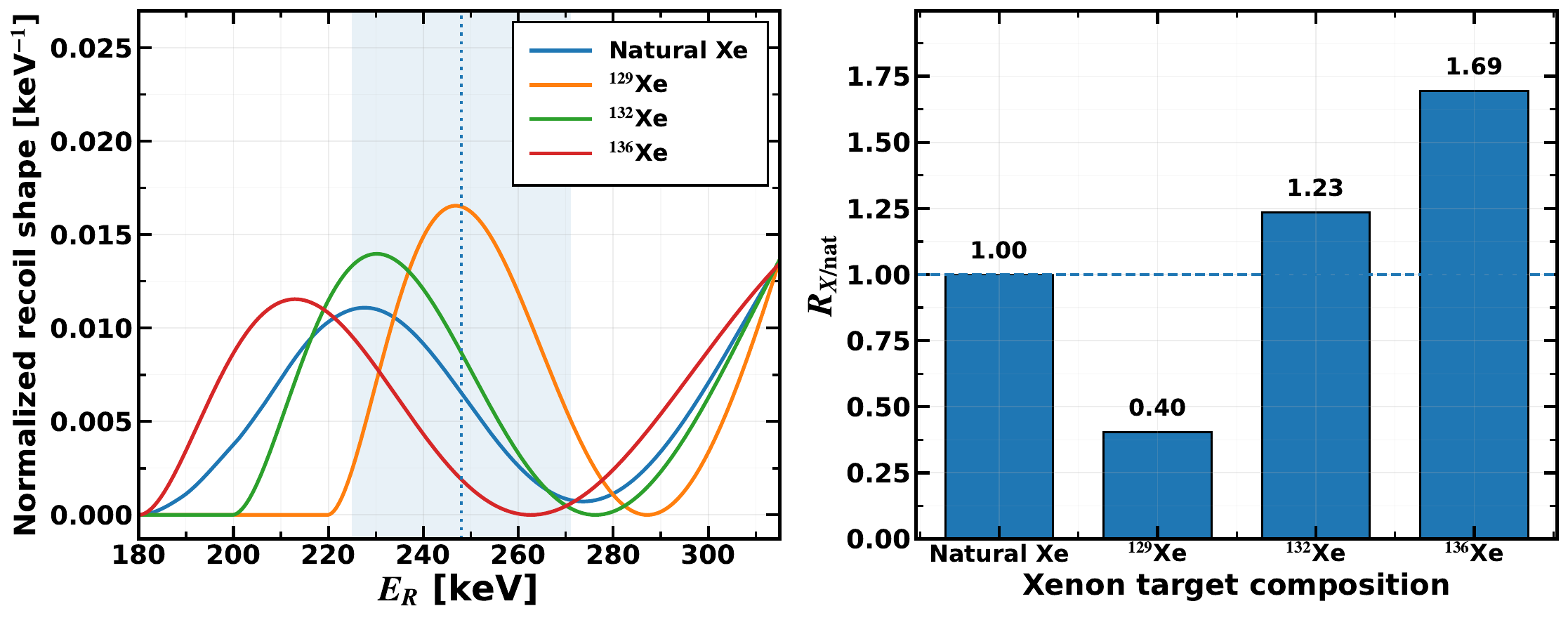}
\caption{\textbf{Natural versus isotope-modified xenon.}
Left: shapes normalized over $180$--$320~\keV$ for natural xenon and idealized pure
$^{129}$Xe, $^{132}$Xe, and $^{136}$Xe targets.
Right: integrated response ratios in $225$--$271~\keV$ at the same microscopic coupling and
detector mass.  These are proof-of-principle composition diagnostics, not forecasts for a
specific enriched detector.}
\label{fig:enriched}
\end{figure*}

For the benchmark we obtain
\begin{equation}
\begin{aligned}
R_{129/{\rm nat}}&=0.404,\qquad
R_{132/{\rm nat}}=1.235,\\
R_{136/{\rm nat}}&=1.693.
\end{aligned}
\label{eq:targetratios}
\end{equation}
Thus changing the isotope mixture modifies both shape and normalization even before detector
effects are included.

\section{Halo dependence and edge amplification}
\label{sec:halo}

The isotope-filtering effect is strongest where the prediction is also most sensitive to the
high-speed halo.  A local linearization gives
\begin{equation}
\delta\ln\eta
\simeq
\frac{\partial\ln\eta}{\partial v_{\min}}\,
\delta v_{\min},
\label{eq:amplification}
\end{equation}
but the edge behavior can be made more explicit.  Define
\begin{equation}
\Delta v
=
v_{\max}-v_{\min}>0.
\end{equation}
For the second branch of the sharply truncated SHM and $\Delta v\to0^+$,
\begin{equation}
\eta(v_{\min})
\simeq
\frac{
z e^{-z^2}
}{
N_{\rm esc}\sqrt{\pi}\,v_Ev_0^2
}
(\Delta v)^2,
\label{eq:etaedge}
\end{equation}
so that
\begin{equation}
\frac{\partial\ln\eta}{\partial v_{\min}}
\simeq
-\frac{2}{\Delta v}.
\label{eq:etaslope}
\end{equation}
Equation~\eqref{eq:amplification} is useful only when the isotope displacement is small
compared with the scale over which $\eta$ varies.  When a shift crosses the cutoff we use the
exact halo integral rather than this linearization.

\begin{figure*}[t]
\centering
\includegraphics[width=0.96\textwidth]{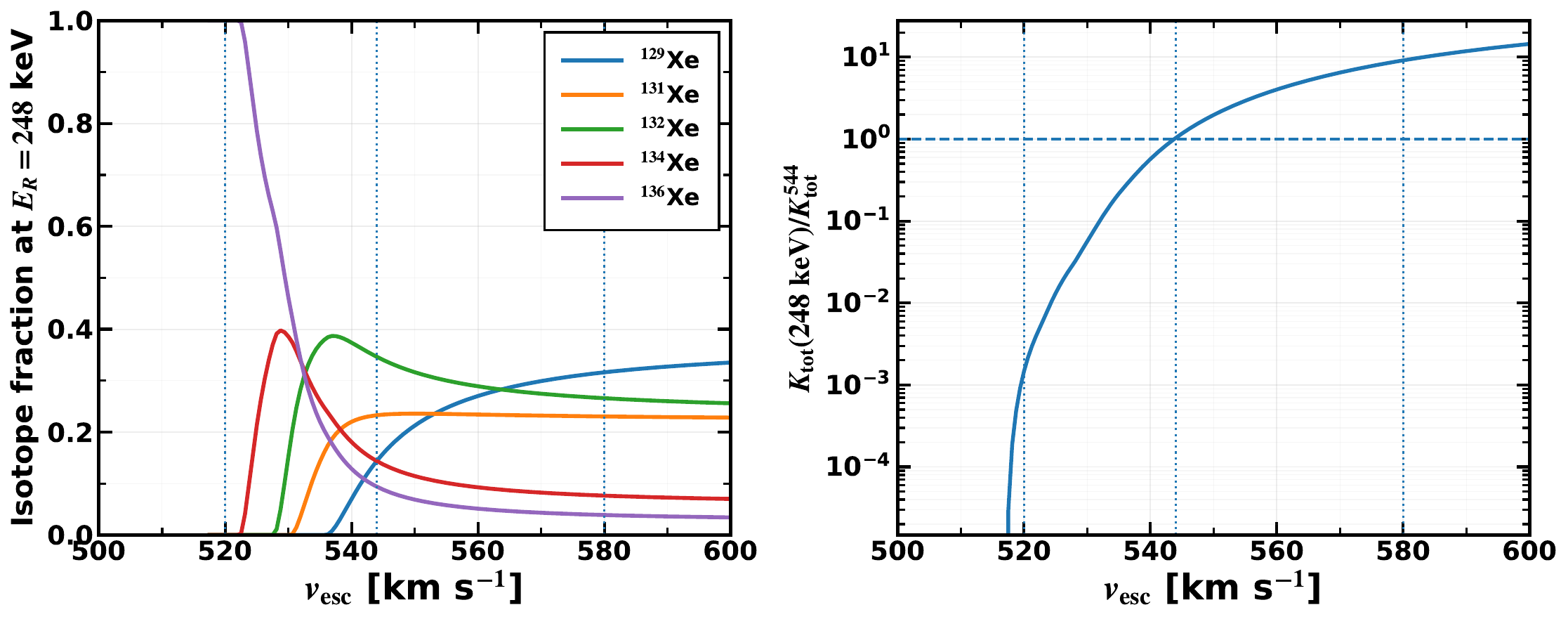}
\caption{\textbf{Halo sensitivity of isotope filtering.}
Left: full Helm-weighted isotope fractions at $E_R=248~\keV$ versus $v_{\rm esc}$.
Right: the total natural-xenon differential kernel at the same recoil, normalized to its
value at $v_{\rm esc}=544~\kms$.
Vertical dotted lines mark $520$, $544$, and $580~\kms$.
The scan is a sensitivity exercise rather than a confidence band.}
\label{fig:halo}
\end{figure*}

Figure~\ref{fig:halo} varies $v_{\rm esc}$ continuously while keeping $v_0$ and $v_E$
fixed.  Undefined fractions below the first-support threshold are masked.

At $v_{\rm esc}=520~\kms$, only $^{136}$Xe has support at $248~\keV$, so its fraction is
exactly one, but the total response is only $1.43\times10^{-3}$ of the baseline value.
At $544~\kms$ the full fractions include
$f_{129}=0.144$, $f_{132}=0.346$, and $f_{136}=0.094$.
At $580~\kms$ these become approximately $0.316$, $0.266$, and $0.0388$, while the total
response is $8.86$ times the baseline.  A large normalized fraction near threshold therefore
must not be confused with a large detectable rate.

The structural statement is more robust than any one numerical fraction: nearby isotope
masses become dynamically important whenever their $v_{\min}$ values sample a rapidly
falling part of a common halo integral.  A full annual prediction, however, requires
$v_E(t)$ and exposure weighting \cite{McCabe2026}.

\section{Nuclear-physics systematics at high recoil}
\label{sec:nuclear}

At $248~\keV$ the xenon momentum transfer is
\begin{equation}
q=\sqrt{2m_AE_R}
\simeq0.24\text{--}0.25~\GeV,
\label{eq:qvalue}
\end{equation}
across the naturally abundant isotopes.  In this regime the Helm response is already strongly
suppressed and varies rapidly with isotope, as quantified by the benchmark values in
Table~\ref{tab:benchmark} and by the contrast between the two panels of
Fig.~\ref{fig:decomposition}.  A separate form-factor figure is therefore unnecessary for the
main result: the left panel of Fig.~\ref{fig:decomposition} isolates the kinematics, while the
right panel shows how the assumed nuclear response reshapes the isotope hierarchy.

The simple replacement
$A^2F_A^2(q)\rightarrow \mathcal W_A(q)$ preserves the same mean-inverse-speed weighting
only for interactions whose differential cross section factorizes as
\begin{equation}
\frac{\dd\sigma_A}{\dd E_R}
\propto
\frac{\mathcal W_A(q)}{v^2}.
\label{eq:factorizing}
\end{equation}
It is not a general prescription for arbitrary nonrelativistic operators.  For example, if
\begin{equation}
\frac{\dd\sigma_A}{\dd E_R}
=
\frac{a_A(E_R)}{v^2}
+
b_A(E_R),
\end{equation}
then the rate contains two halo moments,
\begin{equation}
\frac{\dd R_A}{\dd E_R}
\propto
a_A(E_R)\eta(v_{\min}^A)
+
b_A(E_R)\widetilde\eta(v_{\min}^A),
\end{equation}
with
\begin{equation}
\widetilde\eta(v_{\min})
=
\int_{v>v_{\min}}
\dd^3v\,f_E(\bm v)\,v.
\end{equation}
More general velocity-dependent operators can require still other moments and consistent
inelastic transverse-velocity factors \cite{Fitzpatrick2013,Anand2014}.  The
isotope-dependent support from Sec.~\ref{sec:kinematics} remains, but the relative isotope
weights become operator dependent.

\section{Experimental meaning and detector-level limitations}
\label{sec:experimental}

A conventional xenon time-projection chamber does not identify the isotope that scattered in
an individual recoil.  The fractions $f_A(E_R)$ are therefore latent components of a signal
model, not tagged observables.  The detector-level prediction is schematically
\begin{equation}
\frac{\dd N}{\dd E_{\rm obs}}
=
\mathcal E
\int\dd E_R\,
\epsilon(E_R)\,
G(E_{\rm obs},E_R)
\sum_A
\frac{\dd R_A}{\dd E_R},
\label{eq:detector}
\end{equation}
where $\mathcal E$ is the exposure, $\epsilon(E_R)$ is the efficiency/acceptance, and
$G(E_{\rm obs},E_R)$ is normalized as
\begin{equation}
\int \dd E_{\rm obs}\,
G(E_{\rm obs},E_R)=1.
\end{equation}
This convention keeps acceptance in $\epsilon$ and avoids double counting it in $G$.

At fixed true recoil energy, an isotope-independent efficiency cancels from
Eq.~\eqref{eq:fraction}.  It does not generally cancel from integrated quantities after
smearing and energy-dependent acceptance.  A full LZ recast would therefore sum the
isotope-resolved true-recoil spectra before detector smearing and likelihood evaluation.

The $225$--$271~\keV$ interval used in this paper is a theory illustration centered on
$248~\keV$.  It is not a signal region, a confidence interval, or the complete LZ analysis
window.  The quoted LZ event has separate statistical and systematic energy uncertainties
\cite{LZ2026}.

{\color{black}
\section{Solar-capture consistency and the IceCube bound}
\label{sec:icecube}

The Sun can probe endothermic splittings beyond terrestrial xenon kinematics
because an infalling halo particle is gravitationally accelerated.  If $u$ is
its asymptotic speed and $v_{{\rm esc},\odot}(r)$ is the local solar escape
speed, the speed at radius $r$ is
\begin{equation}
w^2(r,u)=u^2+v_{{\rm esc},\odot}^2(r).
\label{eq:solarspeed}
\end{equation}
Inelastic scattering on a solar nucleus of mass $m_A$ is kinematically
possible only when
\begin{equation}
\delta < \frac{1}{2}\mu_{\chi A}w^2(r,u).
\label{eq:solarcapturekin}
\end{equation}
The much larger speeds available in the solar interior therefore allow capture
for splittings that are already inaccessible to terrestrial xenon.

For a canonical thermal Higgsino near $m_\chi\simeq1.08~\TeV$, the large
off-diagonal electroweak interaction can produce efficient solar capture.
Combining the resulting annihilation signal with IceCube limits on
high-energy neutrinos from the Sun leads to a lower bound on the neutral-state
splitting.  Recent treatments find a characteristic requirement in the range
$\delta\gtrsim0.51$--$0.57~\MeV$, with a representative robust value near
$0.566~\MeV$ once post-capture cooling and equilibration are included
\cite{PospelovRamani2026,DiMauroShaikh2026,IceCube2025}.
Consequently, the benchmark $\delta=365~\keV$ used in the present paper would
not be viable if interpreted as a full-density canonical thermal Higgsino.

The qualification ``if interpreted as a thermal Higgsino'' is essential.
IceCube constrains the neutrino flux produced after solar capture and
annihilation, rather than $\delta$ directly.  A model-independent application
would require, at minimum, the absolute scattering normalization and operator,
the solar capture integral over elemental abundances, the annihilation
spectrum and branching fractions, and the capture--annihilation evolution.
Indeed, endothermic pseudo-Dirac and other interaction structures can evade
the Higgsino mapping even for splittings in the few-hundred-keV range
\cite{DiMauroShaikh2026}.  We therefore do not impose
Eq.~\eqref{eq:icecubecontext} on the generic isotope maps in
Fig.~\ref{fig:dominant}; instead it is used to delimit the applicability of a
specific electroweak realization.
}

{\color{black}
This complementarity is also reflected in other recent analyses.  Independent studies
have used Super-Kamiokande and IceCube data to test Higgsino and more general
inelastic-dark-matter interpretations of the LZ event
\cite{Bose2026,NguyenLindenHooper2026}.  The vector-like-lepton interpretation
of Ref.~\cite{ElahiSchwaller2026} likewise emphasizes the usefulness of solar capture,
whereas cosmic-ray-boosted scenarios can avoid relying on the same Galactic
high-velocity tail that controls endothermic halo scattering
\cite{Heikinheimo2026}.
}

\section{Relation to current LZ interpretations}
\label{sec:relation}

The closest conceptual comparisons help delimit the novelty.  The kinematic-edge analysis of
Ref.~\cite{DiMauro2026} emphasizes the same high-speed-tail regime, but not the
isotope-weight redistribution studied here.  The seasonal analysis of
Ref.~\cite{McCabe2026} shows how strongly an endothermic high-recoil signal can respond to
changes in the laboratory velocity; our Fig.~\ref{fig:halo} resolves part of that sensitivity
into isotope components.  The nuclear-interference study of Ref.~\cite{Khan2026} highlights
a different source of high-$q$ spectral structure.  Finally, XENON1T already established the
practical importance of isotope-resolved inelastic calculations \cite{XENON1T2024}.  The
present contribution is the quantitative organization of these established ingredients into
an edge-amplified isotope-composition diagnostic.

The mechanism is complementary to ultraviolet completions proposed for LZ230616.  Any
endothermic model that places xenon recoils near the halo edge inherits the isotope-dependent
mapping in Eq.~\eqref{eq:vmin}, irrespective of whether the splitting arises from a
pseudo-Dirac mass, an electroweak multiplet, a dark photon, a $B-L$ sector, or an
extra-dimensional construction.  Solar capture, sidebands, annual modulation, and
multi-element targets remain essential complementary tests
\cite{DentNewstead2026,Bose2026,LeeYoun2026,NguyenLindenHooper2026,DiMauroShaikh2026}.

Exothermic and boosted-DM explanations provide useful control cases because they need not
rely on the same Galactic high-speed threshold
\cite{DentNewstead2026,DeLima2026,BaerBarger2026,Alhazmi2026,Kannike2026,Heikinheimo2026}.
A pronounced edge-amplified isotope pattern is therefore characteristic of thresholded halo
scattering, although its detailed magnitude remains interaction dependent.

\section{Compact control checks}
\label{sec:discussion}

Two compact checks are sufficient to show that the main conclusions do not arise from a
purely cosmetic isotope decomposition.  First, replacing natural xenon by a single
representative target with
$A_{\rm eff}=\sum_A\xi_A A=131.388$ and
$m_{\rm eff}=\sum_A\xi_A m_A=122.271~\GeV$
changes the integrated $225$--$271~\keV$ response by about $8\%$ at the benchmark,
\begin{equation}
R_{\rm eff/nat}=1.078.
\end{equation}
The corresponding spectral residual is larger close to the high-$q$ structure.  Thus the full
isotope sum affects the predicted spectrum, not merely an internal bookkeeping convention.

Second, we repeated the kinematic-only calculation at the same dark-matter mass but with
$\delta=200~\keV$, where the xenon isotopes probe the interior of the halo rather than its
upper edge.  In that control case the isotope fractions vary smoothly with recoil energy,
whereas the $\delta=365~\keV$ benchmark shows rapid redistribution as successive isotope
thresholds approach the support boundary.  This verifies that the pronounced filtering is an
edge effect rather than a generic consequence of summing neighboring isotopes.

For all normalized isotope fractions used in the figures we impose the explicit support rule
$\sum_A K_A>0$; points with vanishing total response are masked rather than assigned artificial
fractions.  The numerical implementation also verifies $0\leq f_A\leq1$ and
$\sum_A f_A=1$ wherever the total response is nonzero.  These checks are sufficient for the
focused diagnostic purpose of the present paper without introducing additional figures.

\section{Conclusions}
\label{sec:conclusions}

We have studied a target-composition effect that becomes important when endothermic
dark-matter scattering approaches the upper support boundary of a Galactic halo model.
The effect does not arise because isotope-resolved xenon calculations are new; they are not.
It arises because, near a sharp kinematic edge, percent-level differences in isotope mass can
produce order-unity changes in the relative halo weights.

At the benchmark $m_\chi=1.1~\TeV$, $\delta=365~\keV$, and $E_R=248~\keV$,
the approximate nuclear-mass convention used consistently in our notebook gives
$v_{\min}^{124}=800.77~\kms$ and
$v_{\min}^{136}=767.60~\kms$, compared with
$v_{\max}=794.5~\kms$.  The lightest isotopes are therefore excluded or marginal while the
heavier isotopes sample progressively less suppressed parts of the tail.

Kinematics alone does not determine the final isotope hierarchy.  At the same recoil energy,
the kinematic-only $^{136}$Xe fraction is about $0.289$, but after Helm weighting it is
about $0.094$, while $^{132}$Xe supplies about $0.346$ of the full response.  The
high-$q$ nuclear response therefore competes directly with the threshold ordering.

The composition comparison provides an experimentally meaningful expression of the same
physics.  At equal detector mass and microscopic coupling, the idealized
$^{129}$Xe, $^{132}$Xe, and $^{136}$Xe targets give window-integrated responses of
approximately $0.404$, $1.235$, and $1.693$ relative to natural xenon.  A single
effective-xenon nucleus differs from the full natural-isotope sum by about $8\%$ in the same
window and can distort the spectral shape more strongly.

The precise fractions and dominant-isotope maps remain sensitive to the high-speed halo and
to the nuclear-response model.  We therefore regard the isotope-dependent support boundaries,
the paired kinematic/nuclear comparison, and the failure of a single-mass approximation near
the edge as the most robust outputs.  A detector-level interpretation of LZ230616 would
require efficiency, resolution, sidebands, backgrounds, and a full likelihood analysis and is
outside the scope of this work.

{\color{black}
An additional qualification applies to electroweak Higgsino realizations.
For a full-density thermal Higgsino near $1.1~\TeV$, solar-capture analyses
based on IceCube require a splitting of order $0.51$--$0.57~\MeV$ and hence
exclude the present $365~\keV$ benchmark in that specific realization
\cite{PospelovRamani2026,DiMauroShaikh2026}.  This model-dependent exclusion
does not alter the isotope-filtering mechanism derived here, but it shows that
a viable ultraviolet completion must also satisfy solar-capture and
annihilation constraints.
}

If future xenon experiments accumulate a population of high-energy recoils, isotope
composition can provide an internal consistency test of an endothermic interpretation:
neighboring nuclei with the same atomic number but slightly different masses must map the same
underlying halo into a correlated set of threshold and spectral distortions.
\newpage
\appendix

\section{Analytic mean inverse speed}
\label{app:eta}

Define
\begin{equation}
x=\frac{v_{\min}}{v_0},
\qquad
y=\frac{v_E}{v_0},
\qquad
z=\frac{v_{\rm esc}}{v_0}.
\end{equation}
For the branch structure below we assume $v_{\min}\ge0$ and
$0<v_E<v_{\rm esc}$.  With $N_{\rm esc}$ from Eq.~\eqref{eq:Nesc}, for
$x<z-y$,
\begin{equation}
\eta(v_{\min})
=
\frac{
\erf(x+y)-\erf(x-y)
-\dfrac{4y}{\sqrt{\pi}}e^{-z^2}
}{
2N_{\rm esc}v_E
}.
\label{eq:eta1}
\end{equation}
For $z-y\le x<z+y$,
\begin{equation}
\eta(v_{\min})
=
\frac{
\erf(z)-\erf(x-y)
-\dfrac{2(z+y-x)}{\sqrt{\pi}}e^{-z^2}
}{
2N_{\rm esc}v_E
}.
\label{eq:eta2}
\end{equation}
Finally, for $x\ge z+y$,
\begin{equation}
\eta(v_{\min})=0.
\label{eq:eta3}
\end{equation}
Near $x=z+y$ the two nonzero terms in Eq.~\eqref{eq:eta2} nearly cancel.  The numerical
notebook therefore evaluates the final part of this branch with the positive series implied by
Eq.~\eqref{eq:etaedge}.

\section{Natural xenon inputs}
\label{app:isotopes}

Table~\ref{tab:isotopes} lists the machine-readable inputs used by the final notebook.
Neutral-atom masses and natural atom fractions are from NIST \cite{NISTXe}; the approximate
nuclear masses follow Eq.~\eqref{eq:massconv}.

\begin{table}[h]
\centering
\caption{\textbf{Natural xenon inputs used in the calculation.}}
\label{tab:isotopes}
\resizebox{\columnwidth}{!}{
\begin{tabular}{cccc}
\toprule
Isotope &
$\xi_A$ &
$m_A^{\rm atom}$ [u] &
$m_A$ [GeV]\\
\midrule
$^{124}$Xe & 0.000952 & 123.9058920 & 115.390014\\
$^{126}$Xe & 0.000890 & 125.9042983 & 117.251517\\
$^{128}$Xe & 0.019102 & 127.9035310 & 119.113791\\
$^{129}$Xe & 0.264006 & 128.9047808611 & 120.046449\\
$^{130}$Xe & 0.040710 & 129.903509349 & 120.976759\\
$^{131}$Xe & 0.212324 & 130.90508406 & 121.909720\\
$^{132}$Xe & 0.269086 & 131.9041550856 & 122.840349\\
$^{134}$Xe & 0.104357 & 133.90539466 & 124.704491\\
$^{136}$Xe & 0.088573 & 135.907214484 & 126.569175\\
\bottomrule
\end{tabular}}
\end{table}


\begin{thebibliography}{99}

\bibitem{LZ2026}
D. S. Akerib \textit{et al.} (LUX-ZEPLIN Collaboration),
``Search for dark matter particle interactions in an extended nuclear recoil energy window with
the LUX-ZEPLIN (LZ) experiment,''
arXiv:2609.02823 [hep-ex] (2026).

\bibitem{TuckerSmith2001}
D. Tucker-Smith and N. Weiner,
Phys. Rev. D \textbf{64}, 043502 (2001),
arXiv:hep-ph/0101138.

\bibitem{TuckerSmith2005}
D. Tucker-Smith and N. Weiner,
Phys. Rev. D \textbf{72}, 063509 (2005),
arXiv:hep-ph/0402065.

\bibitem{Su2026}
L. Su, J. M. Yang, and W.-N. Yang,
``Inelastic Dark Matter Signature at High Recoil Energy in LUX-ZEPLIN and CRESST,''
arXiv:2609.01475 [hep-ph] (2026).

\bibitem{DiMauro2026}
M. Di Mauro,
``Dark Matter at the Kinematic Edge: Interpreting the 248 keV LZ Nuclear-Recoil Candidate,''
arXiv:2609.02608 [hep-ph] (2026).

\bibitem{McCabe2026}
C. McCabe,
``Seasonal dark matter from the LUX-ZEPLIN high-energy event,''
arXiv:2609.04181 [hep-ph] (2026).

\bibitem{Yang2026}
M. Yang, Q.-f. Wu, Y.-L. S. Tsai, and Y.-Z. Fan,
``Multi-Messenger and Paleo-Detector Probes of the LZ Dark Matter Signal,''
arXiv:2609.06640 [hep-ph] (2026).

\bibitem{Fan2026}
Z.-T. Fan, H.-J. He, Y.-C. Wang, and Y. Zhao,
``Inelastic Dark Matter and High-Energy Recoil Signatures in LZ,''
arXiv:2609.10491 [hep-ph] (2026).

\bibitem{Visinelli2026}
L. Visinelli,
``A Peccei--Quinn Origin for Inelastic Electroweak Dark Matter after
LUX-ZEPLIN,''
arXiv:2609.02807 [hep-ph] (2026).

\bibitem{Freese2026}
K. Freese and D. P. Theodosopoulos,
``Higgsino Dark Matter Interpretation of the LUX-ZEPLIN 248 keV Nuclear-Recoil Event,''
arXiv:2609.01583 [hep-ph] (2026).

\bibitem{Cheung2026}
K. Cheung, S. K. Kang, and R. Kumar,
``From LUX-ZEPLIN to Colliders: Probing Higgsino Dark Matter,''
arXiv:2609.08712 [hep-ph] (2026).

\bibitem{Bisal2026}
S. Bisal, J. Cao, and F. Li,
``Higgsino Dark Matter Interpretation of the LZ High-Recoil Event in the GNMSSM with TeV-Scale
Gauginos,''
arXiv:2609.07811 [hep-ph] (2026).

\bibitem{WangXiao2026}
L. Wang and Y. Xiao,
``The Inert Doublet Model of Dark Matter and the LUX-ZEPLIN High-Recoil Event,''
arXiv:2609.06571 [hep-ph] (2026).

\bibitem{Borah2026}
D. Borah, S. K. Sahoo, N. Sahu, and S. Sharma,
``Inelastic Singlet-Doublet Fermion Dark Matter in light of the 248 keV LZ event,''
arXiv:2609.07800 [hep-ph] (2026).

\bibitem{OkadaSeto2026}
N. Okada and O. Seto,
``Inelastic $B-L$ scalar dark matter and the LUX-ZEPLIN event,''
arXiv:2609.06909 [hep-ph] (2026).

\bibitem{KumarPrajapati2026}
R. Kumar and H. K. Prajapati,
``Generalized Chiral $U(1)_{B-L}$ with Inelastic Scalar Dark Matter for the LZ 248 keV Event,''
arXiv:2609.10827 [hep-ph] (2026).

\bibitem{QiSun2026}
X. Qi and H. Sun,
``Nonthermal Solar Stalling of an Inelastic Scalar Signal in Xenon,''
arXiv:2609.10636 [hep-ph] (2026).

\bibitem{Yamashita2026}
K. Yamashita,
``Inelastic Dark Photon Dark Matter for the LUX-ZEPLIN High-Recoil Event and the Galactic Halo
Gamma-Ray Excess,''
arXiv:2609.02868 [hep-ph] (2026).

\bibitem{Zhu2026}
P. Zhu \textit{et al.},
``Endothermic dark matter with a light dark photon and the LUX--ZEPLIN high-energy nuclear-recoil
candidate,''
arXiv:2609.09015 [hep-ph] (2026).

\bibitem{Yuan2026}
G.-W. Yuan, B. Zhang, W.-Y. Cao, L. Feng, and R. Yang,
``ALP-mediated inelastic dark matter and the LUX-ZEPLIN high-recoil candidate event LZ230616,''
arXiv:2609.08893 [hep-ph] (2026).

\bibitem{He2026}
Y. He,
``Transition magnetic-dipole dark matter and the LZ230616 high-recoil candidate,''
arXiv:2609.10453 [hep-ph] (2026).

\bibitem{AhmedLeontaris2026}
W. Ahmed and G. K. Leontaris,
``A Dark-Dimension Origin of Geometric Inelastic Dark Matter:
The LUX-ZEPLIN High-Recoil Event and Multi-Target Tests,''
arXiv:2609.07138 [hep-ph] (2026).

\bibitem{LeeRandall2026}
V. S. H. Lee and L. Randall,
``A Warped Extra Dimensional Candidate for the LZ 248 keV Event,''
arXiv:2609.09136 [hep-ph] (2026).

\bibitem{DentNewstead2026}
J. B. Dent and J. L. Newstead,
``Exothermic and Endothermic Inelastic Dark Matter Interpretations at LZ:
Sideband Constraints and Future Prospects,''
arXiv:2609.04673 [hep-ph] (2026).

\bibitem{DeLima2026}
C. H. de Lima,
``Exothermic Dark Matter at LZ,''
arXiv:2609.05204 [hep-ph] (2026).

\bibitem{BaerBarger2026}
H. Baer and V. Barger,
``Exothermic dark matter and the 248 keV nuclear recoil in LUX-ZEPLIN,''
arXiv:2609.06153 [hep-ph] (2026).

\bibitem{LouLu2026}
Y. Lou and S.-T. Lu,
``Fermionic Dark Matter Absorption and the High-Energy Event in LUX-ZEPLIN,''
arXiv:2609.01592 [hep-ph] (2026).

\bibitem{JeesunMajumdar2026}
S. Jeesun and A. Majumdar,
``Atmospheric neutrino up-scattering explanation of LZ 2026 excess,''
arXiv:2609.04185 [hep-ph] (2026).

\bibitem{Alhazmi2026}
H. Alhazmi, D. Kim, K. Kong, J.-C. Park, and S. Shin,
``High-Energy Nuclear Recoils from Boosted Dark Matter for the LZ 248-keV Event:
Beyond the Halo-Dependent High-Velocity Tail,''
arXiv:2609.06890 [hep-ph] (2026).

\bibitem{Kannike2026}
K. Kannike, M. Raidal, and A. Strumia,
``Boosted dark particles and the LZ nuclear recoil event,''
arXiv:2609.07742 [hep-ph] (2026).

\bibitem{Heikinheimo2026}
M. Heikinheimo and N. Zimmermann,
``Cosmic ray boosted dark matter with momentum dependent interactions can explain the LZ 248 keV
event,''
arXiv:2609.11600 [hep-ph] (2026).

\bibitem{Liang2026}
J.-H. Liang, Z. Liu, V. Q. Tran, and Y. Xu,
``LZ Nuclear-Recoil Excess from Boosted Light Magnetic Dipole-dipole Dark Matter,''
arXiv:2609.06756 [hep-ph] (2026).

\bibitem{AghaieStrumia2026}
M. Aghaie and A. Strumia,
``Neutron disappearance and the LZ nuclear recoil event,''
arXiv:2609.09037 [hep-ph] (2026).

\bibitem{Khan2026}
I. Khan, S. Capozziello, G. Mustafa, F. Atamurotov, A. Abdujabbarov, and C. Yuan,
``Nuclear interference versus dark sector excitation in the 248 keV LUX-ZEPLIN recoil candidate,''
arXiv:2609.09230 [hep-ph] (2026).

\bibitem{Unwin2026}
J. Unwin,
``Axion Portal Dark Matter and the LUX-ZEPLIN High-Recoil Event,''
arXiv:2609.04186 [hep-ph] (2026).

\bibitem{Das2026}
P. Das, B. Karmakar, S. Mahapatra, and P. K. Paul,
``Inelastic Self-interacting Dark Matter and LUX-ZEPLIN 248 keV Event in a Dirac Modular Inverse
Seesaw,''
arXiv:2609.06825 [hep-ph] (2026).

\bibitem{Chattaraj2026}
A. Chattaraj, A. Majumdar, D. K. Papoulias, and R. Srivastava,
``Can Elastic Neutrino Scattering Account for the LZ230616 Event?,''
arXiv:2609.10504 [hep-ph] (2026).

\bibitem{ElahiSchwaller2026}
F. Elahi and P. Schwaller,
``A Vector-Like Lepton Interpretation of the High-Energy Nuclear Recoil Candidate
in LUX-ZEPLIN,''
arXiv:2609.08993 [hep-ph] (2026).

\bibitem{XENON1T2024}
E. Aprile \textit{et al.} (XENON Collaboration),
``Effective field theory and inelastic dark matter results from XENON1T,''
Phys. Rev. D \textbf{109}, 112017 (2024),
arXiv:2210.07591 [hep-ex].

\bibitem{PospelovRamani2026}
M. Pospelov and H. Ramani,
``Strong Constraints on Higgsino Dark Matter from Solar Capture,''
arXiv:2609.02775 [hep-ph] (2026).

\bibitem{DiMauroShaikh2026}
M. Di Mauro and H. Shaikh,
``Solar Capture Tests of Inelastic Dark Matter after the LZ High-Recoil Event,''
arXiv:2609.06760 [hep-ph] (2026).

\bibitem{IceCube2025}
R. Abbasi \textit{et al.} (IceCube Collaboration),
``Search for High-Energy Neutrinos From the Sun Using Ten Years of IceCube Data,''
arXiv:2507.08457 [hep-ex] (2025).

\bibitem{LewinSmith1996}
J. D. Lewin and P. F. Smith,
Astropart. Phys. \textbf{6}, 87 (1996).

\bibitem{Fitzpatrick2013}
A. L. Fitzpatrick, W. Haxton, E. Katz, N. Lubbers, and Y. Xu,
JCAP \textbf{02}, 004 (2013),
arXiv:1203.3542 [hep-ph].

\bibitem{Anand2014}
N. Anand, A. L. Fitzpatrick, and W. C. Haxton,
Phys. Rev. C \textbf{89}, 065501 (2014),
arXiv:1308.6288 [hep-ph].

\bibitem{NISTXe}
National Institute of Standards and Technology,
``Atomic Weights and Isotopic Compositions for Xenon,''
NIST Physical Measurement Laboratory,
\url{https://physics.nist.gov/cgi-bin/Compositions/stand_alone.pl?ascii=ascii\&ele=Xe}
(accessed 12 September 2026).

\bibitem{Bose2026}
D. Bose \textit{et al.},
``Not so good $\nu$s for Higgsino dark matter as LZ excess:
stringent limits from Super-Kamiokande and IceCube,''
arXiv:2609.07807 [hep-ph] (2026).

\bibitem{NguyenLindenHooper2026}
T. T. Q. Nguyen, T. Linden, and D. Hooper,
``Solar Neutrino Constraints on Inelastic Dark Matter Scattering in Light of Recent LUX-ZEPLIN
Observations,''
arXiv:2609.11833 [hep-ph] (2026).

\bibitem{LeeYoun2026}
S. J. Lee and T. Youn,
``Mixing-suppressed inelastic dark matter: a minimal model for the LZ 248 keV event,''
arXiv:2609.09138 [hep-ph] (2026).

\end{thebibliography}
\end{document}